# Unraveling the Unconventional Order of a High-Mobility Indacenodithiophene-Benzothiadiazole Copolymer

Camila Cendra[†], Luke Balhorn[†], Weimin Zhang[‡], Kathryn O'Hara[‖], Karsten Bruening[§], Christopher J. Tassone[§], Hans-Georg Steinrück[§,††], Mengning Liang[§], Michael F. Toney[§, ‡‡], Iain McCulloch[‡, #], Michael L. Chabinyc[‖], Alberto Salleo[†]*, Christopher J. Takacs[§]*

*E-mail: asalleo@stanford.edu, ctakacs@slac.stanford.edu

[†]Department of Materials Science and Engineering, Stanford University, Stanford, California 94305, United States

[‡]Physical Science and Engineering Division KAUST Solar Center (KSC), King Abdullah University of Science and Technology (KAUST), Thuwal, Saudi Arabia

[‖]Materials Department, University of California Santa Barbara, Santa Barbara, California 93106, United States

[§]Stanford Synchrotron Radiation Lightsource, SLAC National Accelerator Laboratory, Menlo Park, California 94025, United States

[#] Department of Chemistry, University of Oxford, Oxford OX1 3TA, United Kingdom

[††] Department Chemie, Universität Paderborn, 33098 Paderborn, Germany

[‡‡] Department of Chemical and Biological Engineering, University of Colorado Boulder, Boulder, Colorado 80303, United States

**Abstract**

A new class of donor-acceptor (D-A) copolymers found to produce high charge carrier mobilities competitive with amorphous silicon (>1 $cm^2V^{-1}s^{-1}$) exhibits the puzzling microstructure of substantial local order, however lacking long-range order and crystallinity previously deemed necessary for achieving high mobility. Here, we demonstrate the application of low-dose transmission electron microscopy to image and quantify the nanoscale and mesoscale organization of an archetypal D-A copolymer across areas comparable to electronic devices (≈ 9 $\mu m^2$). The local structure is spatially resolved by mapping the backbone (001) spacing reflection, revealing nanocrystallites of aligned polymer chains throughout nearly the entire film. Analysis of the nanoscale structure of its ordered domains suggests significant short- and medium-range order and preferential grain boundary orientations. Moreover, we provide insights into the rich, interconnected mesoscale organization of this new family of D-A copolymers by analysis of the local orientational spatial autocorrelations.

Due to their solution-processability, flexibility, and synthetic tunability, conjugated polymers (CPs) have emerged as a promising materials family for solar cells,[1] flexible displays,[2] energy storage,[3] chemical sensors,[4] neural probes,[5] and neuromorphic computing.[6] In all of these applications, functionality and performance depend strongly on the solid-state molecular and mesoscale organization.[7–9] Charge transport requires continuous electronic coupling across all length scales (i.e., atomic to mesoscale) and consideration of anisotropies at the molecular level.[10,11] There are notable examples of higher order mesoscale organization and molecular packing structures in polymeric systems, but their discovery is often serendipitous, and their roles are poorly understood.[12–27] A grand challenge in this area of materials science is to uncover how molecular structure and processing produce the hierarchal solid-state structures that ultimately govern materials properties, be they optoelectronic, electrochemical, or even mechanical. Indeed, a detailed understanding of the morphology has remained elusive in CPs, largely due to the difficulty to capture intricate, complex, and heterogeneous structures using typical materials science characterization approaches such as electron microscopy and X-ray scattering techniques. There is a critical need for the development of methods capable of unraveling and understanding the exceptional and unexpected aspects of self-assembly in materials like CPs across length scales.

Several well-performing donor-acceptor (D-A) copolymers described as structurally "disordered" owing to lack of conventional long-range crystalline order, exhibit large persistence lengths.[28–30] This extended structure along the backbone direction and efficient charge transport suggest short-to-medium range nanometer-scale structural organization are likely both present and important. Assessing the extent of ordering at short- and medium-range length scales (i.e., nearest neighbors and next-nearest neighbors and beyond, respectively) in these materials is imperative towards a complete picture of structural heterogeneity. Low-dose transmission electron microscopy (TEM) has proven to be a viable approach to characterize the local and intercrystallite order of CP thin-films from the atomic to mesoscopic scales.[14,15,37–40,24,26,31–36] In glassy materials, TEM techniques have revealed nanometer-scale ordered clusters in a more disordered matrix.[40–43] Beyond direct damage, electron-beam/materials interactions have several other practical consequences on TEM image contrast. These include contrast reduction through damage-induced mechanical strain/vibrations and charge-fluctuations within the sample that blur image contrast.[44] With the introduction of active-pixel counting electron detectors, dose-fractionated TEM image stacks (movies) under low-dose conditions offer the potential to address and correct these contrast issues to improve imaging resolution and signal to noise.[45]

Here, we present an approach to low-dose high-resolution TEM (HRTEM) that takes advantage of the many instrumental advances developed to study biological specimens[46] to reveal previously undetected structural features. Specifically, we detect a surprising degree of short- and medium-range order in a high-performance indacenodithiophene-co-benzothiadiazole copolymer (IDT-BT) whose microstructure is commonly referred as "amorphous-like" based on its broad, weak X-ray diffraction.[47] Our work showcases the potential of new TEM techniques and analysis methods to advance our understanding of complex, functional soft matter. Analysis reveals the presence of locally aligned regions of polymer chains throughout the entirety of the sample with frequent overlaps of the domains. We leverage the vast amount of available information to describe and quantify higher-order structural relationships across various length scales. First, we identify and segment self-similar regions exhibiting alignment to estimate the size distribution of nanoscale domains (mean value of $\approx$ 15 nm) and examine the nature of short- and medium-range order. Second, we examine the rich, interconnected mesoscale organization of IDT-BT through estimation of the local orientational correlations. The results agree with complementary synchrotron and free-electron laser hard X-ray characterization. Our results exemplify the applicability of using low-dose HRTEM imaging and cryo-EM inspired analysis methods to characterize the nano- and mesoscale (≈ 9 $\mu m^2$) organization of IDT-BT, which is essential for further understanding the role of structure in a new class of high-performing CPs lacking pronounced long-range order.

## Results and Discussion

Initially described by Zhang et al.,[48] IDT-BT (Figure 1a) is a prominent example of a new class of donor-acceptor (D-A) copolymers lacking pronounced long-range order,[29,47] yet displaying remarkable optoelectronic properties and high charge carrier mobilities ($\mu \approx$ 2–10 $cm^2V^{-1}s^{-1}$).[49,50] These properties were attributed to nearly planar, torsion-free backbone conformations[51] and low degrees of energetic disorder,[49] with recent evidence suggesting the formation of excitons at close-crossing points where chains aggregate strongly.[52,53]

The alkyl sidechains, which render CPs solution-processable, influence the solid-state macromolecular order and play a critical role in the self-assembly process.[54] In IDT-BT, the long alkyl sidechains are likely disordered, creating regions of excluded volume around the planar backbones that can extend above and below the rigid plane of the backbone. Moreover, at the level of a single IDT unit, there is a local inversion symmetry where the sidechains attach, that could create significant steric hindrances.

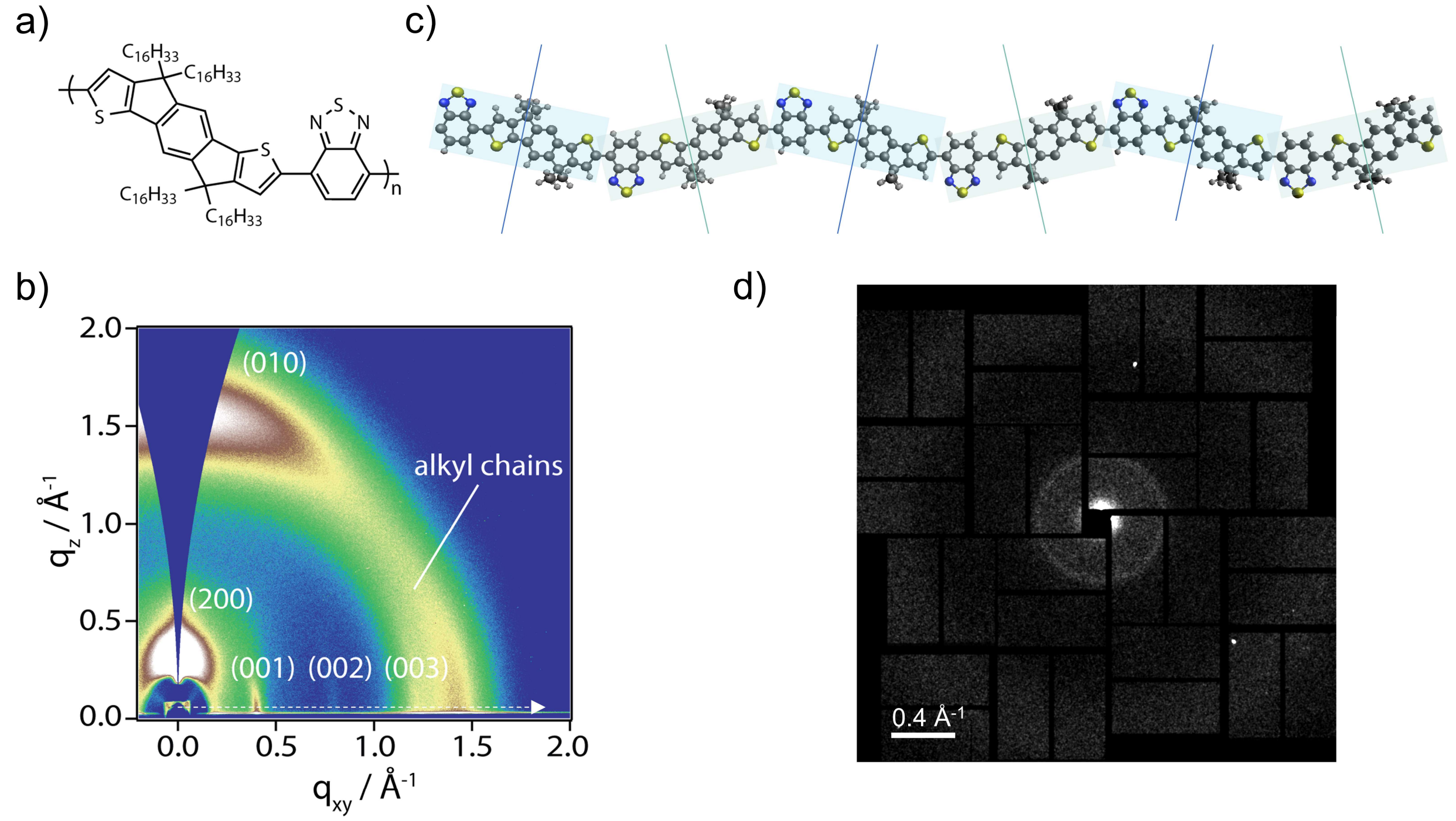

**Figure 1.** Molecular structure and packing properties of IDT-BT. **a)** Chemical structure. **b)** Optimized gas-phase geometry with sidechains omitted. Blue and green rectangular boxes and lines denote the orientation of the monomer relative to the molecular axis. **c)** 2D GIWAXS pattern. Dashed arrow denotes in-plane scattering direction. **d)** Single-shot FEL X-ray diffraction pattern (in transmission mode) using a nanofocus beam. The estimated beam size is ≈ 250 nm at full width at half maximum.

The backbone is thought to be both highly planar and linear (Figure S1, Supporting Information).[49,51,55] Importantly, simple rotational defects along the chain have unusual consequences since the monomer entrance and exit angles are approximately co-linear and offset (Figure 1b). As a result, in thin-films, the molecular structure of IDT-BT along the backbone direction is likely to present various conformational isomers that maintain a rigid, planar rod topology. Thus, the chain direction may wander but is not expected to bend the chain strongly, resulting in an entropically linear/rod-like chain.[56] Combined with the sterically encumbered and unusual local symmetry of the IDT unit in the monomer, we expect a range of complex and potentially unconventional morphologies may be present in solid-state.

Synchrotron and free-electron laser (FEL) hard X-ray scattering experiments were used to investigate thin-film packing characteristics averaged over millimeter and sub-micron scale, respectively. The bulk packing characteristics of IDT-BT as probed by grazing-incidence wide-angle X-ray scattering (GIWAXS) agree with literature[47,48] (Figure 1c and Figure S2, Supporting Information), where the most prominent feature is a single family of sharp in-plane ($q_{xy}$) peaks at $\approx 0.4$, 0.8, and 1.2 $Å^{-1}$ assigned to the infrequently observed "backbone" reflections (00l), a feature associated with the polymer repeat unit length (d-spacing $\approx$ 1.56 nm). The series of narrow (00l) peaks is accompanied by a broad halo at q $\approx$ 1.2–1.5 $Å^{-1}$ attributed to diffraction from disordered sidechains and broad diffraction alkyl (200) and π-stacking (010) peaks from crystallites with seemingly edge-on and face-on populations, respectively.[47,49] Something of a misnomer, these "backbone" reflections also require precise intermolecular shifts of adjacent chains. These directions are generally thought to be "soft" since local tilting/slipping of neighboring chains can disrupt translational order. Given the prevalence of these "backbone" features, the associated "alkyl-stacking" and "π-stacking" features[47] are comparatively weak and broad, reflecting the short coherence lengths and reduced crystallinity along these intermolecular directions (Table S1, Supporting Information).

Interestingly, IDTBT also appears capable of forming different, highly ordered morphologies at the interface compared to those suggested by hard X-ray experiments. Polarized near-edge X-ray spectroscopy conducted by Zhang et al. indicate a strong degree of molecular orientation near the surface, observing a nearly-perfect face-on texture.[47] Nevertheless, the orientation measured by the weak X-ray diffraction suggests both edge-on and face-on populations in the bulk (Figure 1c), differing from the spectroscopy results.[47] Despite the apparently conflicting thin-film microstructure of high molecular order while lacking long-range order and crystallinity, scanning tunneling microscopy (STM) measurements[55] on sub-monolayer films show that IDT-BT is capable, under certain processing conditions, of efficient side chain interdigitation, even when the side chain-bearing units have different orientations, and that local strands are capable of aligning parallel to each other. Furthermore, molecular dynamics (MD) simulations of the unit cell of IDT-BT predict side chain interdigitation of adjacent lamellae.[49] These results suggest that traditional bulk X-ray diffraction or spectroscopy techniques fail to fully capture the nano- to mesoscale structure of materials lacking long-range order such as IDT-BT.

To further investigate mesoscale heterogeneity, we conducted FEL X-ray diffraction measurements at the Linac Coherent Light Source (Figure 1d). The quasi-in-plane diffraction of the sample is acquired in normal incidence using a "diffraction before destruction" methodology[57] to avoid X-ray damage. A single, high-intensity X-ray pulse ($\approx$ 100 fs) is focused to $\approx$ 250 nm FWHM and the diffraction is captured on a segmented 2D area detector. Each diffraction pattern of IDT-BT shows a sharp ring at q $\approx$ 0.4 $Å^{-1}$ and a broad halo at q $\approx$ 1.2–1.5 $Å^{-1}$, matching well the observed features in GIWAXS (Figure 1c). The backbone ring (001) is mostly uniform with slight variations suggestive of a finite but large number of crystallites contributing without discernable local orientational order. There is some scattered intensity at q-values below the backbone (001) peak; however, azimuthal integration of the data (Figure S3, Supporting Information) does not show any feature resembling a peak that could be attributed to in-plane alkyl stacking of the polymer. The data suggest that the microstructure of IDT-BT must be probed at length scales significantly smaller than the size of the 250 nm X-ray beam in order to enable assessment of the mesoscale heterogeneity.[58,59] Towards this end, we visualize the structure of IDT-BT with nanometer-scale resolution using low-dose cryogenic HRTEM.

HRTEM is a direct imaging method capable of extracting high-fidelity, local structural information that can spatially resolve diffraction from features probed by X-ray techniques. With the electron beam at normal incidence, HRTEM can directly image the in-plane backbone diffraction,[26,32] which corresponds to the $q_{xy}$ slice of the GIWAXS pattern (Figure 1c) and the sharp ring in the FEL data (Figure 1d). A schematic of the HRTEM imaging setup is shown in Figure 2a, where the imaging conditions (further described in Section 1 and 3, Supporting Information) are optimized to image the backbone lattice spacing of IDT-BT. In agreement with X-ray results, the computed power spectrum over a region of ≈ 200 nm in diameter (Figure 2a) shows a sharp diffraction ring related to the backbone repeat unit, indicating the presence of aligned backbone sections with overall isotropic orientation. A representative 100 nm x 100 nm region of IDT-BT, its computed power spectrum, and the computed power spectra from smaller subregions are shown in Figure 2b and Figure 2c. Whereas the spectrum of the full region shows a continuous ring from the backbone at $q \approx 0.4$ Å$^{-1}$, the spectra of the individual 12 nm x 12 nm sub-regions exhibit discrete, well-defined peaks. Within these subregions, the backbone peak is spotty, suggesting relatively well-defined nanocrystalline domains that often overlap within the small windows. It is important to note that while it is not possible to determine the order of the polymer along the beam direction (through the thickness of the film, ≈ 40 nm), given that we frequently observe overlapping structures, a number of layers are presumed to exist through the thickness of the film. Moreover, we are only probing those nanocrystalline domains that have components that satisfy the Bragg condition, and some crystalline regions may not be imaged in this analysis, suggesting the degree of order is still underestimated.

To better visualize the local alignment of IDT-BT and show the coupling of local chain aggregates from the nanoscale to mesoscale, we use digital image processing methods (Section 3 and Section 4, Supporting Information) and resolve the local average molecular orientation over length scales ranging from nanometers to tens or hundreds of nanometers. We start by visualizing a small 50 nm x 50 nm image region of IDT-BT. The director field (Figure 2d) represents a spatial mapping of the average direction of the polymer backbones visible in the projected image, and it is composed of short lines drawn parallel to the backbone direction extracted from the computed power spectrum over small 12 nm x 12 nm regions of the film.[32] Brandley et al. have applied similar methods to map the orientation of carbon nanotubes from scanning electron micrographs.[60] We observe nanometer-scale regions of locally aligned polymer chains, and frequent overlapping between regions. Using the local value of the molecular director as the tangent vector, lines representing the predominant orientation of the backbone are propagated to follow the slow contours of the director field map, as shown in Figure 2e. The reconstructed map of the polymer backbone orientation displays the presence of nanoscale domains of aligned polymer chains and overlapping of domains. Similar features are observed through a reconstruction of the average polymer orientation along the backbone direction over larger length scales (Figure 3 and Figure S4, Supporting Information). Whereas at short length scales ($\lessapprox$ 10 nm), the polymer is likely to be locally aligned and exhibit a well-defined orientation, over larger length scales (hundreds of nanometers), these regions of locally aligned backbones show an overall isotropic orientation.

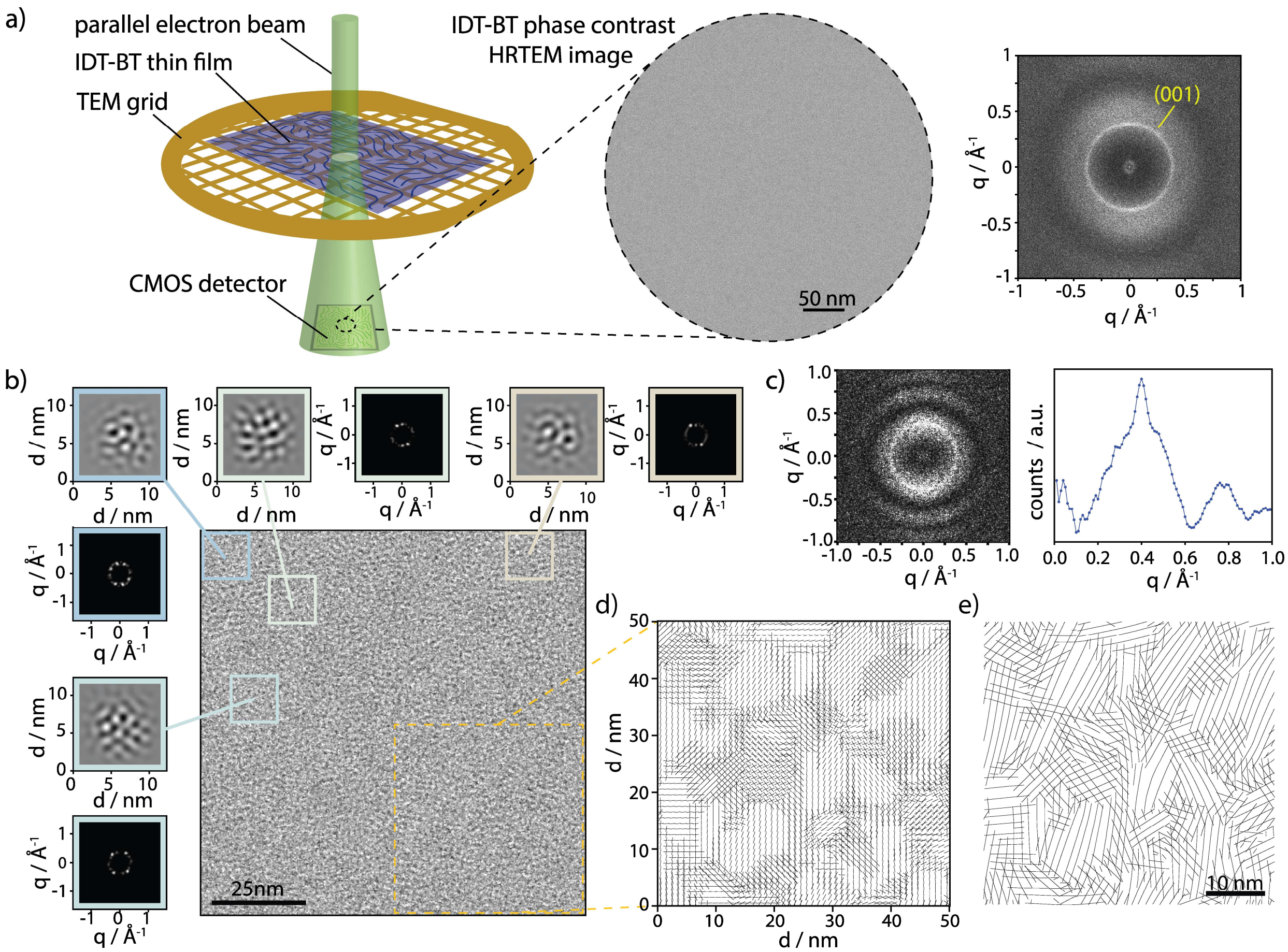

**Figure 2.** Visualization of nanoscale ordering along the backbone direction in IDT-BT. **a)** Schematic of the HRTEM imaging process (left), a phase contrast HRTEM image of a 0.125 $\mu m^2$ region of IDT-BT (middle) and its power spectrum (right). The power spectrum shows an isotropic ring at a spatial frequency of $q \approx 0.4$ Å$^{-1}$ corresponding to the (001) reflection of the backbone (i.e., along the chain). This power spectrum is nearly-equivalent to the nanodiffraction patterns presented in Figure 1d. **b)** Exploratory visualization of nanoscale features on a representative 100 nm x 100 nm region of IDT-BT. Expanded views of four bandpass-filtered 12.3 nm x 12.3 nm regions illustrate the polymer backbone arrangement and the presence of locally well-defined backbone stacking peaks. Bandpass filter is centered at $q \approx 0.4$Å$^{-1} \pm 0.03$ Å$^{-1}$. **c)** Power spectrum (left) of the HRTEM image shown in part b) and a lineout of an azimuthally integrated power spectrum (right). **d)** Computed director field of the bottom right region of the image shown in part b). **e)** Reconstruction of the predominant polymer orientation extracted from the computed director field. In the director field map, short lines are drawn parallel to the direction of the observed backbone stacking peaks. These short lines are then propagated following the local slope taken from the director field to reconstruct the predominant polymer orientation map.

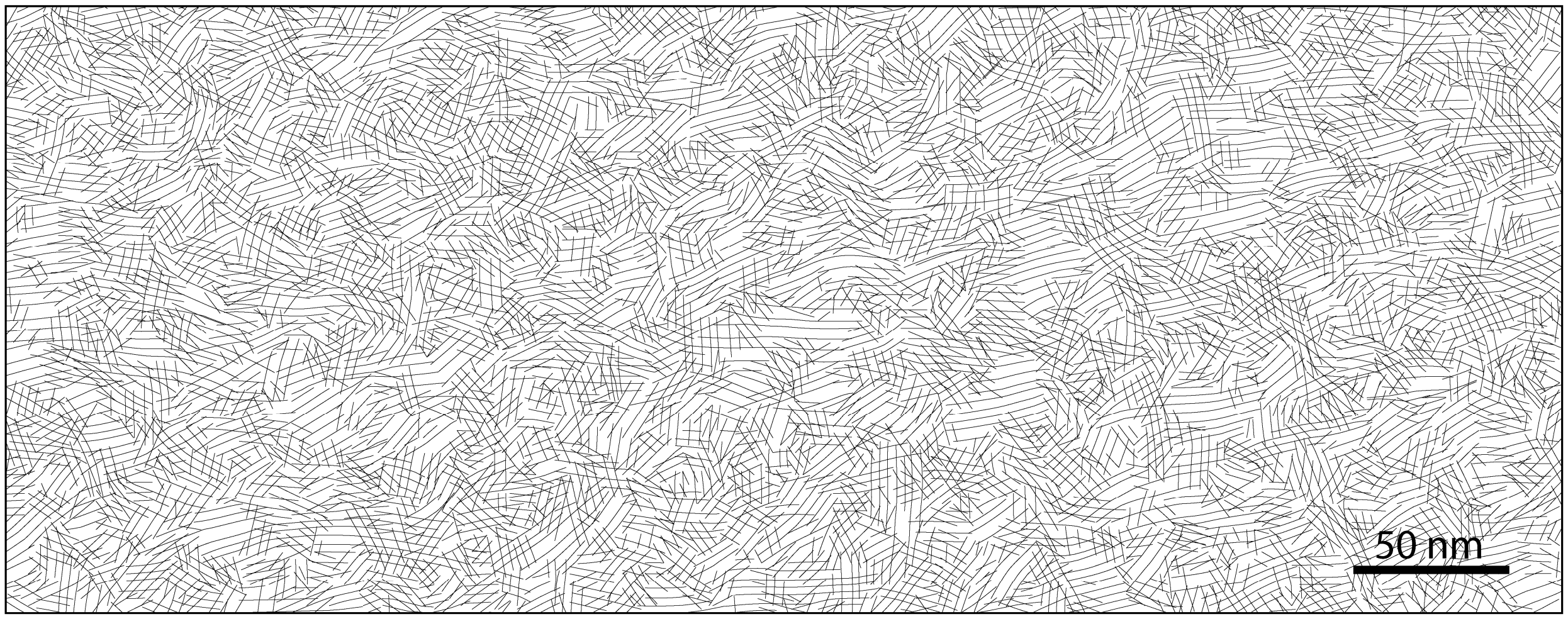

**Figure 3.** Reconstruction of the predominant backbone orientation based on a HRTEM image of IDT-BT. Small (≈ 10 – 20 nm) domains of aligned polymer chains are observed along with a highly overlapping structure.

Gathering statistical information about structural features such as domain size and grain boundaries is of interest for the interpretation of charge transport properties in conjugated polymers.[8,21] To characterize and quantify the extent of the nanoscale domains present in IDT-BT, we develop an algorithm to segment HRTEM images into neighboring regions with similar polymer alignment (within ± 5°) and we identify them as single domains. This choice in angle is chosen to approximately match the angular distribution of the locally averaged domains (*vide infra*) and an informative qualitative picture consistent with the level of disorder. As shown in Figure 4a, this method allows for the visual identification of domains while accounting for spatial overlap between domains. To account for the fact that larger domains will cover larger areas of the sample, we also calculated the cumulative area-weighted distribution (Figure 4b). For half of the sampled area, the domains are $\gtrapprox$ 20 nm. The mean value of the area-weighted distribution (≈ 16 nm) is in agreement with the estimated coherence length (≈ 14 nm) from GIWAXS of the (001) peak using the Scherrer equation (Figure S8 and Table S1, Supporting Information, respectively). The statistical distribution of the grain boundary angles is shown in Figure 4c. The analysis was performed by calculating the relative misorientation between one domain and neighboring or overlapping domains. All grain boundary orientations are observed, with a preference for ≈ 20° and a slighter one for ≈ 90° grain boundaries. Further details are provided in Section 5, Supporting Information.

The domain analysis and segmentation process also offers new opportunities to examine order in macroscopically isotropic samples. Stiff, rod-like polymers can have both crystalline and liquid-crystalline-like phases.[21,61–63] This liquid-crystalline-like character is already apparent from the orientation maps; however, there may be additional weak and/or broad molecular features available. To investigate whether such features exist for the expected in-plane alkyl feature of face-on crystallites, we reorient the calculated power spectra of individually segmented domains to a common backbone orientation and compute the median (Figure S5, Supporting Information). No significant diffraction or diffuse scattering signal is observed in the expected range of an alkyl peak (i.e., between 2–3 nm and approximately orthogonal to the backbone reflection).[47,49] The absence of an in-plane alkyl stacking diffraction peak in the X-ray results (Figure 1), as well as in the reoriented median power spectrum of the domains (Figure S5, Supporting

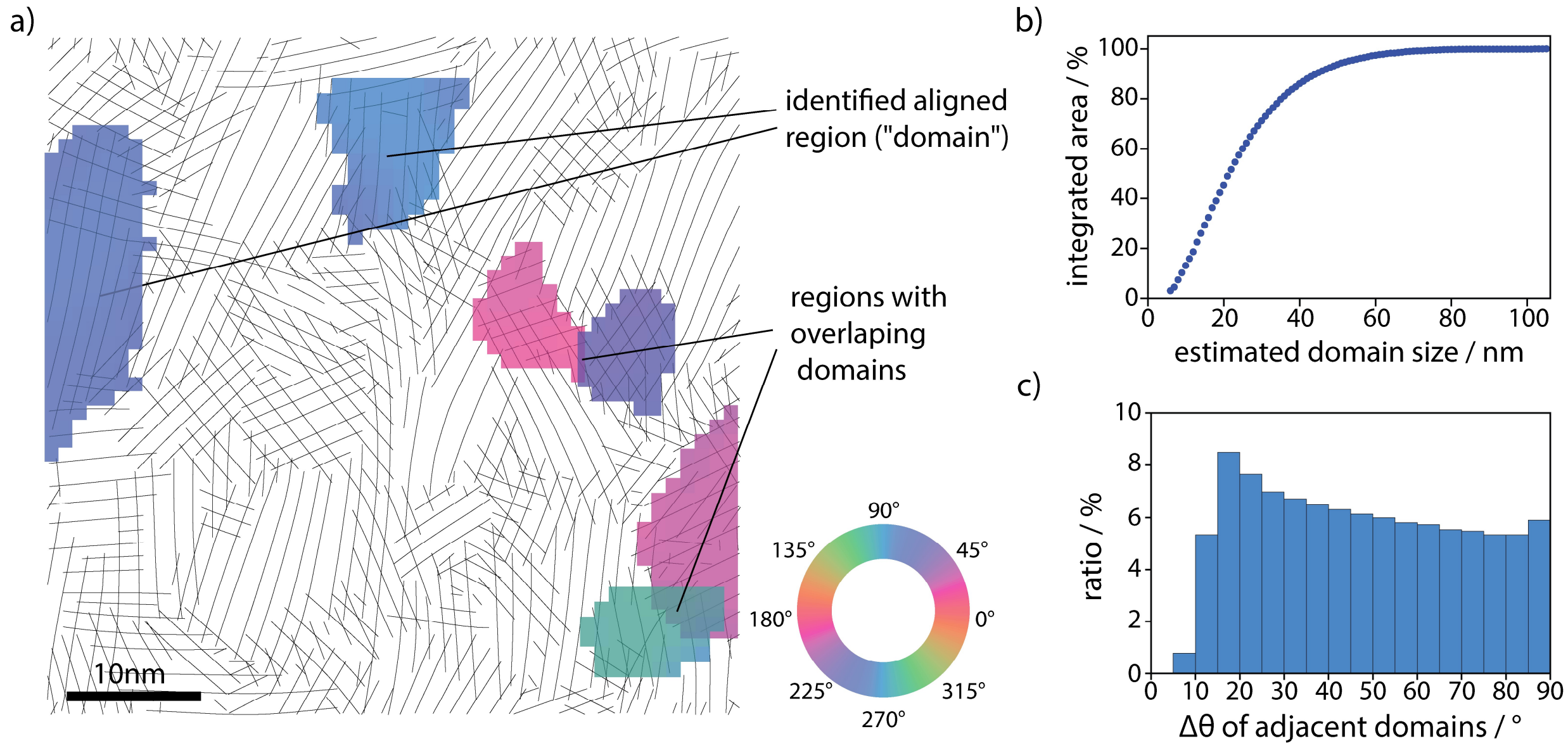

**Figure 4.** Domain identification and analysis. **a)** Illustrative example of the clustering algorithm, which identifies and spatially segments a HRTEM image into regions ("domains") of aligned backbone within ± 5°. The color of each domain is calculated from the median orientation of the identified domains. The map of identified domains is overlaid with the predominant polymer orientation map shown in Figure 2e for illustratory purposes. **b)** Cummulatively area-weighted distribution of domain sizes calculated across various HRTEM images, equivalent to a probed area of ≈ 9 μm$^2$. **c)** Histogram showing the distribution of grain boundary misorientations.

Information) is interesting with several possible explanations. Whereas it is possible than in some regions of the film there exists face-on regions with sidechain interdigitation and the formation of compact assemblies,[55] we have yet to observe these in spin-coated thin-films, suggesting these events are rare or exist primarily at the interface. If the face-on assumption of the planar polymer backbone is correct, it is possible that IDT-BT adopts an altogether different and unknown thin-film structure. Alternatively, the domains may consist of edge-on crystallites.

Given that charge transport in conjugated polymers requires electronic connectivity across various length scales,[11] it is of interest to characterize the extended structural correlations in IDT-BT beyond the size of its domains. We choose a statistical measure that describes spatial relationships of how locally aligned domains change as a function of distance (Δd) and misorientation (Δθ). The probability of finding two locally aligned regions of the polymer separated by a distance Δd and misorientation Δθ follows the autocorrelation function:

$$C(\Delta d, \Delta\theta) = \frac{\sum f(d,\theta) \cdot f(d + \Delta d, \theta + \Delta\theta)|_{d,\theta}}{\sum f(d) \cdot f(d + \Delta d)|_{d}}$$

where $f(d,\theta)$ is a Boolean value (0 or 1) indicating whether at (d, θ) a backbone peak is present. Similarly, $f(d + \Delta d, \theta + \Delta\theta)$ is a Boolean indicating whether there exists a backbone peak at a (Δd, Δθ) relative to the point defined in $f(d,\theta)$. The denominator term, $\sum f(d) \cdot f(d + \Delta d)|_{d}$, is the sum of backbone peaks at separated Δd regardless of their misorientation. This normalization term accounts for the fact that some

distances are more likely to occur than others on the analysis grid. To compare the observed structural correlations in IDT-BT relative to a sample not expected to present clustered regions of locally aligned backbones, the autocorrelations of a simulated sample with random backbone orientations were calculated (representative lineout shown in Figure 5b). Comparisons with IDT-BT are further discussed in Section 6, Supporting Information.

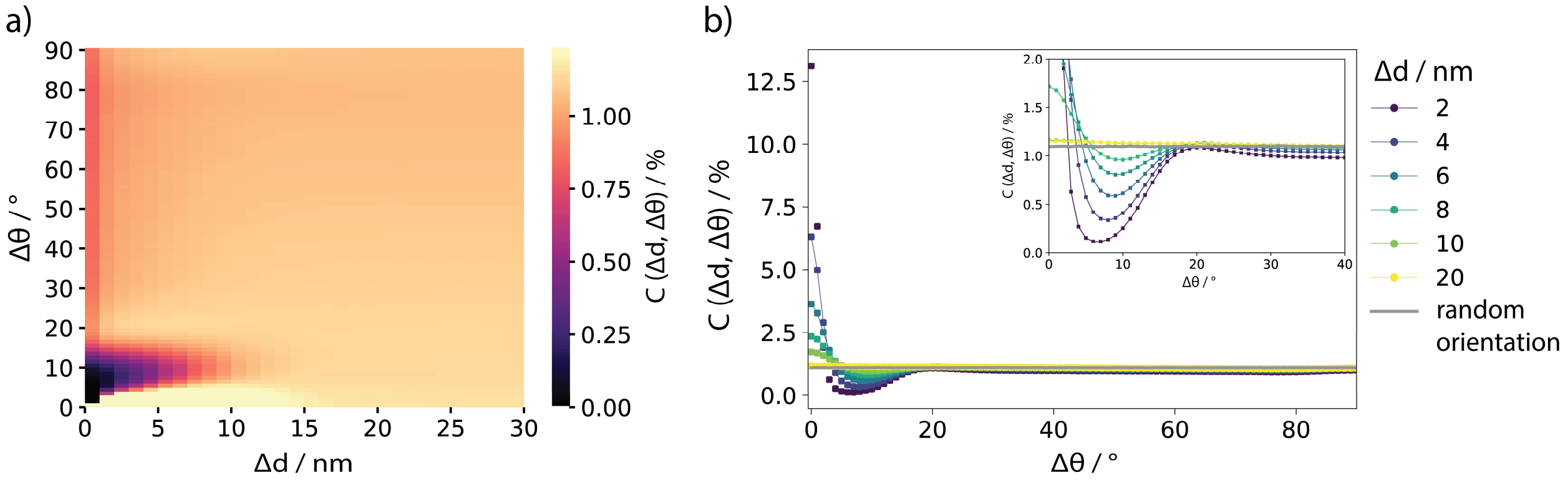

**Figure 5.** Local orientational spatial autocorrelations analysis over a probed area of ≈ 9 μm$^2$ of IDT-BT. **a)** Heatmap of the local orientational autocorrelations, $C(\Delta d, \Delta\theta)$. The scaling range is computed using robust quantiles of $C(\Delta d, \Delta\theta)$ instead of the extreme values. **b)** Extracted linecuts of $C(\Delta d, \Delta\theta)$ at selected distances distances ($\Delta d$). A representative linecut (at $\Delta d$ = 20 nm) for a simulated sample with random backbone orientations is shown in gray. Extreme values were accounted for scaling of the y-axis.

The heatmap of $C(\Delta d, \Delta\theta)$ and extracted linecuts for representative distances (Figure 5) highlight several structural features of interest. For $\Delta d \lessapprox 10$ nm, we observe a prevalence for local alignment of IDT-BT within 5°, which is well in agreement with the median value of the domain size distribution (Section 5, Supporting Information). For $\Delta d \gtrapprox 15$ nm, the autocorrelations of IDT-BT agree with the observed $C(\Delta d, \Delta\theta)$ values of the simulated sample with random orientations (Figure S9, Supporting Information), indicating that correlations do not typically survive beyond ≈ 15 nm and the material exhibits an isotropic behavior at these length scales. Importantly, for $\Delta d \lessapprox 10$ nm there is a lower than random probability for regions of aligned polymer chains to exhibit angular misorientations between ≈ 5–18°, suggesting an impediment for forming low angle grain boundaries. Reminiscent of the observed preferential grain boundaries at ≈ 20° and 90° (Figure 4c), the heatmap z-score of the autocorrelations (Figure S10c,d, Supporting Information) suggests increased likelihood for domain overlapping at these angles for $\Delta d \lessapprox 10$ nm. Despite their weak signal, these features could have profound, non-linear implications in domain interconnectivity.[24,64] We hypothesize these correlations originate from two possible structures. Given that HRTEM images represent a projection of the sample structure through the thickness of the film and that IDT-BT exhibits weak order along the $\pi$-stacking direction (Figure 1), it is possible that IDT-BT nanodomains are not tolerant to defects between layers and that attempts to stack result in a relative rotation between layers that likely maintain a common backbone reflection. Alternatively, defects along the chain axis that maintain backbone planarity, such as a rotation between the IDT and BT units, produce a curvature bend that locally changes the direction of the backbone.

## Conclusion

We have characterized the mesoscale structure with nanometer-scale resolution of a solution-processed thin-film of IDT-BT by leveraging advances in low-dose cryogenic HRTEM coupled with data analytics. Importantly, we have unveiled the nanoscale and mesoscale structure of a material often considered "amorphous-like", finding evidence for unconventional packing structures and remarkable short- and medium-range order. By mapping the (001) backbone spacing reflection, we have identified nanometer-scale regions of locally aligned polymer chains (domains) yet found no evidence for alkyl stacking in these ordered phases, suggesting that solution-processed films of IDT-BT exhibit aspects of liquid-crystalline behavior. Analysis of the local orientational spatial autocorrelations suggests that IDT-BT forms packing structures of domains at preferential overlapping angles, which might affect electronic transport properties. Charge transport in conjugated polymers depends on molecular structure, chain conformation, local aggregation, and mesoscale microstructure. While chain rigidity is an important factor in limiting energetic disorder,[49] it also likely contributes to the development of the nanoscale and mesoscale order we observe by HRTEM. The coexistence of charge delocalization along the rigid backbone favored by the molecular structure, and an advantageous mesostructure featuring medium-range order are instrumental in enabling the outstanding charge transport properties of IDT-BT.

This study showcases the applicability of using HRTEM to characterize the rich, complex mesoscale structure of conjugated polymers with unprecedented detail, shedding light on the pitfalls of using conventional descriptors of order and bulk characterization techniques to elucidate the complex microstructure of materials lacking long-range order and overt crystallinity yet exhibiting outstanding molecular order.

## Acknowledgements

We thank Dr. Elizabeth Montabana, Dr. David Buschnell, and Dr. Dong-Hua Chen for technical assistance with the Tecnai F20 transmission electron microscope. We also thank Sébastien Boutet (LCLS) for beamline assistance. Work by CC and LB on image analysis was supported by the U.S. Department of Energy (DOE), Office of Science, Basic Energy Sciences (BES) under Award DE-SC0020046. CC and AS also gratefully acknowledge financial support from the National Science Foundation Award # DMR 1808401. CJT, KO, and MLC also acknowledge financial support from the National Science Foundation Award # DMR 1436263. We also gratefully acknowledge the support of NVIDIA Corporation with the donation of the Titan V GPU used for this research. Some of this work was performed at the Stanford-SLAC Cryo-EM Facilities, supported by Stanford University, SLAC and the National Institutes of Health S10 Instrumentation Programs. Measurements at Stanford Synchrotron Radiation Lightsource, SLAC National Accelerator Laboratory, were supported by the U.S. Department of Energy, Office of Science, Office of Basic Energy Sciences under Contract No. DE-AC02-76SF00515. Use of the Linac Coherent Light Source (LCLS), SLAC National Accelerator Laboratory, is supported by the U.S. Department of Energy, Office of Science, Office of Basic Energy Sciences under Contract No. DE-AC02-76SF00515.

# Supporting Information

# Unraveling the Unconventional Order of a High-Mobility Indacenodithiophene-Benzothiadiazole Copolymer

Camila Cendra[†], Luke Balhorn[†], Weimin Zhang[‡], Kathryn O'Hara[‖], Karsten Bruening[§], Christopher J. Tassone[§], Hans-Georg Steinrück[§,††], Mengning Liang[§], Michael F. Toney[§, ‡‡], Iain McCulloch[‡, #], Michael L. Chabinyc[‖], Alberto Salleo[†]*, Christopher J. Takacs[§]*

*E-mail: asalleo@stanford.edu, ctakacs@slac.stanford.edu

**Table of Contents**

## 1 Experimental methods

**TEM sample preparation.** Thin-films of IDT-BT (Mn ≈ 109 $kgmol^{-1}$, Mw ≈ 290 $kgmol^{-1}$) for TEM characterization were prepared inside a nitrogen glovebox (< 1ppm $O_2$) by spin coating a solution of IDT-BT dissolved in chloroform (5 mg/mL) and left stirring at 60℃ overnight onto silicon/$SiO_2$ substrates at 1500 rpm for 60 s. Spin coating was followed by drying in $N_2$ at room temperature (1 h) to remove residual solvent. The resulting films were ≈ 40 nm thick. The thin-films were delaminated onto an air-DI water interface by dissolving the $SiO_2$ layer in a 0.5% HF solution and transferred to an ultrathin carbon TEM grid (Ted Pella #01824) for HRTEM imaging.

**HRTEM imaging.** HRTEM images were acquired on a Tecnai F20 transmission electron microscope operating at 200 kV with a direct electron detection camera (Gatan K2) under cryogenic temperatures. The pixel size of the images is 1.9 Å $pixel^{-1}$, and the images have 3838 x 3710 pixels (738 nm x 714 nm). The sample was cooled with liquid $N_2$ to mitigate beam damage during image acquisition. The incident electron beam illuminating a thin-film of IDT-BT is set to be approximately parallel. A small defocus (≈ - 400 nm) is applied to maximize the spatial response around the backbone peak, taking care to avoid the first zero in the contrast transfer function. After moving to an undamaged portion of the sample, the instrument is given time to stabilize mechanically before applying a defocus and exposure. A stack of 48 dose-fractionated images (movies) of 25 ms of exposure for each frame is recorded. The dose rate was 4 to 5 $e^{-}$ $Å^{-2}$ $s^{-1}$, so that each frame represents an electron dose of 1 to 1.25 $e^{-}$ $Å^{-2}$. Image blurring due to specimen-holder drift was monitored between image frames and corrected using reported methods[45,65,66] (more information in Supporting Information Section 4). The total accumulated dosage during the imaging process was restricted to 10 – 15 $e^{-}Å^{-2}$ and fractionated over 8 – 12 frames. Under these electron dosage conditions, there is no apparent damage leading to signal degradation over the spatial frequencies of interest. Following similar methods to Mastronarde et al.[67] and Takacs et al.,[38] SerialEM was used to electronically shift the beam off the optical axis to an undamaged region of the sample and acquire non-overlapping dose-fractionated image stacks in a hexagonal pattern with the beam blanked at all other times.

**GIWAXS characterization.** IDT-BT thin-films for X-ray characterization were prepared by spin coating following identical conditions to the samples prepared for TEM (except for the delamination step). GIWAXS was performed at the Stanford Synchrotron Radiation Lightsource (SSRL) on beamline 11-3 using an area detector (Rayonix MAR-225) and incident energy of 12.73 keV. The distance between sample and detector was 315.65 mm and it was calibrated using a LaB6 polycrystalline standard. The incidence angle (0.1°) was slightly larger than the critical angle, ensuring that we sampled the full depth of the film. The raw data were reduced and analyzed using a combination of Nika 1D SAXS[68] and WAXStools[69] software packages in Igor Pro. GIWAXS measurements were performed in a Helium environment to minimize air scattering and beam damage to samples.

**Free-electron laser X-ray diffraction.** Single-shot diffraction measurements were conducted using the Linac Coherent Light Source (LCLS) at the CXI beamline[70] with the nanofocus chamber. The beam energy was 7996 eV and the CSPAD segmented area detector was used with an approximate sample-to-detector distance of 240 mm. Thin-film samples of IDT-BT were suspended over fine copper mesh and measured in transmission at normal incidence with 51 μm steps between sample positions.

# 2 Supporting figures

## 2.1 Density Functional Theory

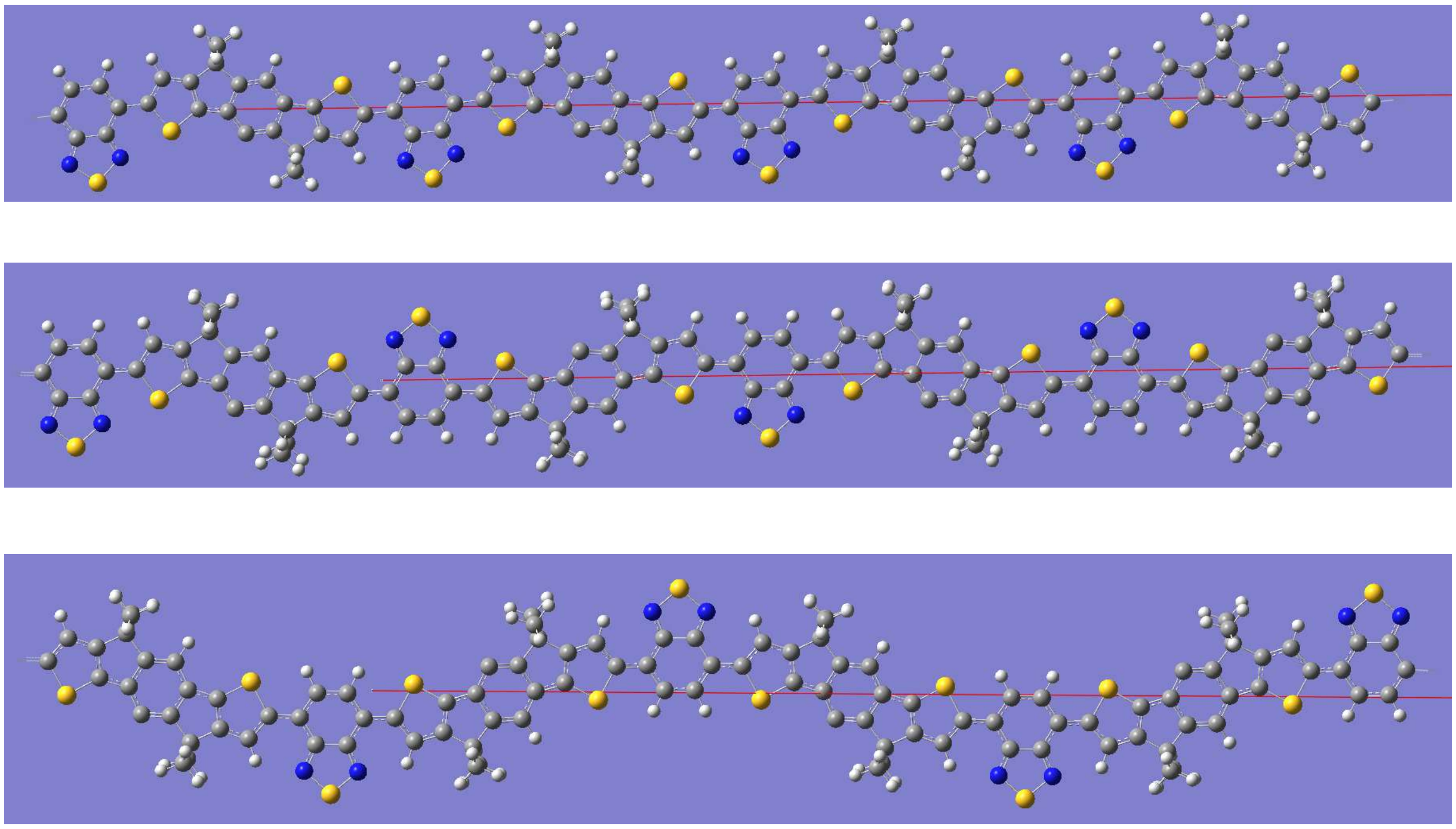

**Figure S1.**Optimized gas-phase geometry at a CAM B3LYP/6-311g level of theory for multiple conformational isomers in IDT-BT. The alkyl side chain groups were replaced by a methyl group, a substitution that does not significantly impact the calculated geometry.[71]

## 2.2 Grazing Incidence wide-angle X-ray scattering

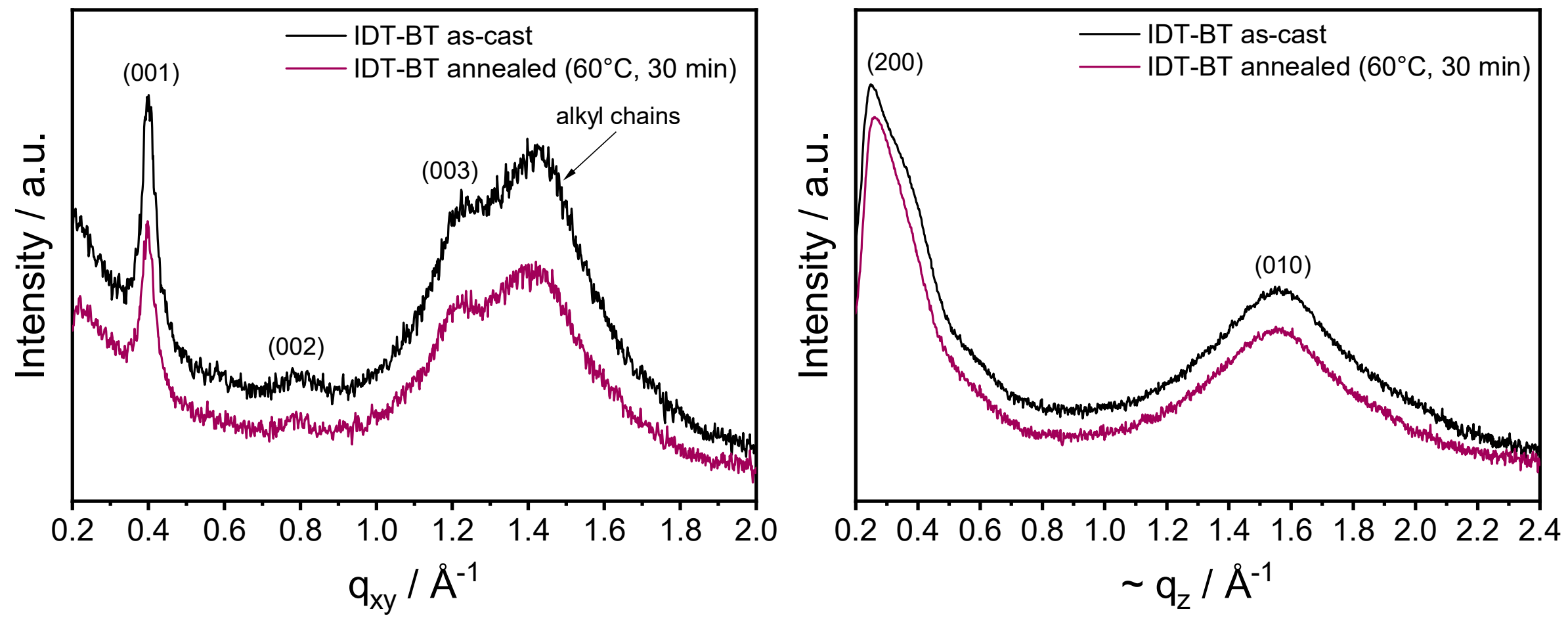

**Figure S2.** In-plane ($q_{xy}$) and out-of-plane ($q_z$) X-ray scattering lineouts for IDT-BT thin-films in their as-cast and annealed (60°C for 30 minutes) condition.

**Table S1.** Grazing-incidence X-ray diffraction peak fits for as-cast and annealed thin-film of IDT-BT. The coherence length along the backbone (001) and π-stacking directions is estimated using the Scherrer equation. Fits for the backbone and π-stacking peaks are extracted from the in-plane (IP) and out-of-plane (OOP) X-ray diffraction lineouts, respectively, shown in Figure S2. The fit of the OOP alkyl stacking peak (200) is not reported since part of this peak is blocked by the beam stop.

| **Sample** | **Peak** | **q / Å$^{-1}$** | **Δq / Å$^{-1}$** | **d / Å** | **Lc / nm** | **# planes** |
|---|---|---|---|---|---|---|
| as-cast IDT-BT | (001) IP | 0.40 | 0.04 | 15.5 | 14.3 | 9.2 |
| | (010) OOP | 1.56 | 0.27 | 4.0 | 2.3 | 5.7 |
| annealed IDT-BT | (001) IP | 0.40 | 0.04 | 15.5 | 14.3 | 9.2 |
| | (010) OOP | 1.55 | 0.28 | 4.1 | 2.3 | 5.6 |

## 2.3 Free-electron laser X-ray diffraction

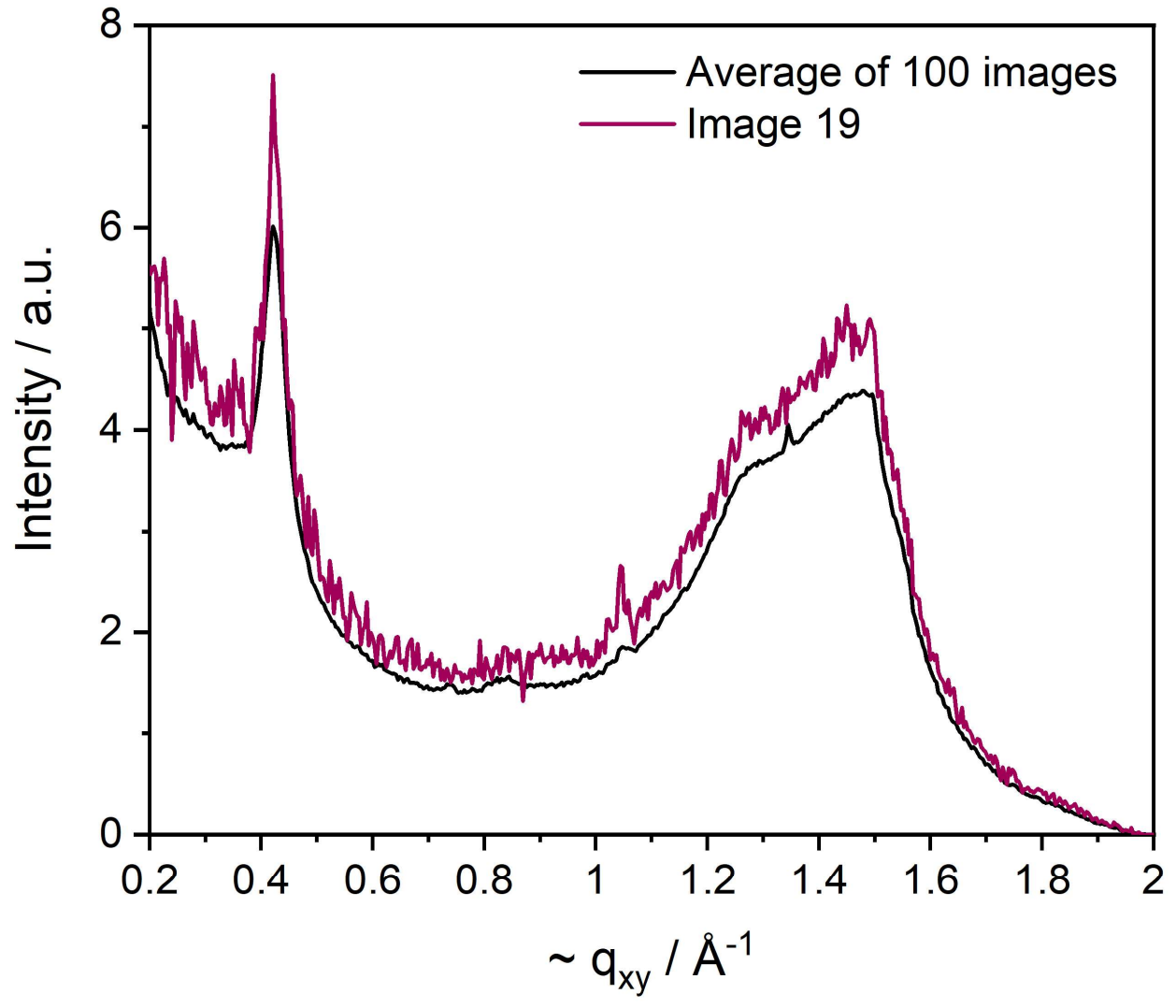

**Figure S3.** Azimuthally averaged free-electron laser diffraction data. A single-shot diffraction pattern (Image 19 of 100 was chosen at random) and the average of the 100 single-shot patterns are shown. We note that due to the curvature of the Edwald's sphere at this energy, the $q_{xy}$ will have a $q_z$ component, particularly at higher q.

### 2.4 Reconstruction of IDT-BT average backbone orientation

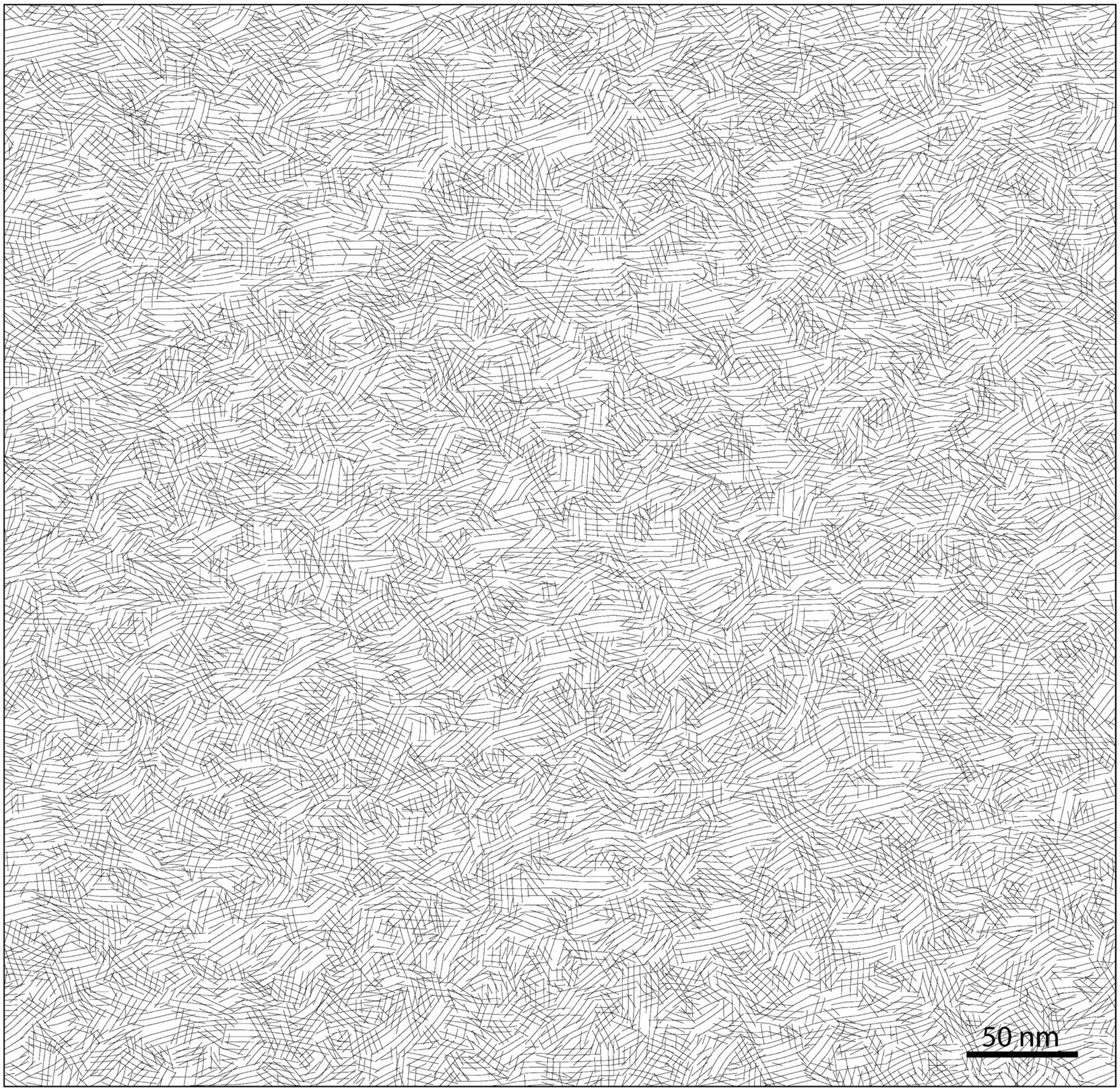

**Figure S4.** Reconstruction of the predominant backbone orientation based on a 500 nm x 500 nm HRTEM image of IDT-BT. Small rigid ordered domains are observed along with a highly overlapping structure.

## 2.5 Median Power Spectrum extracted for identified domains

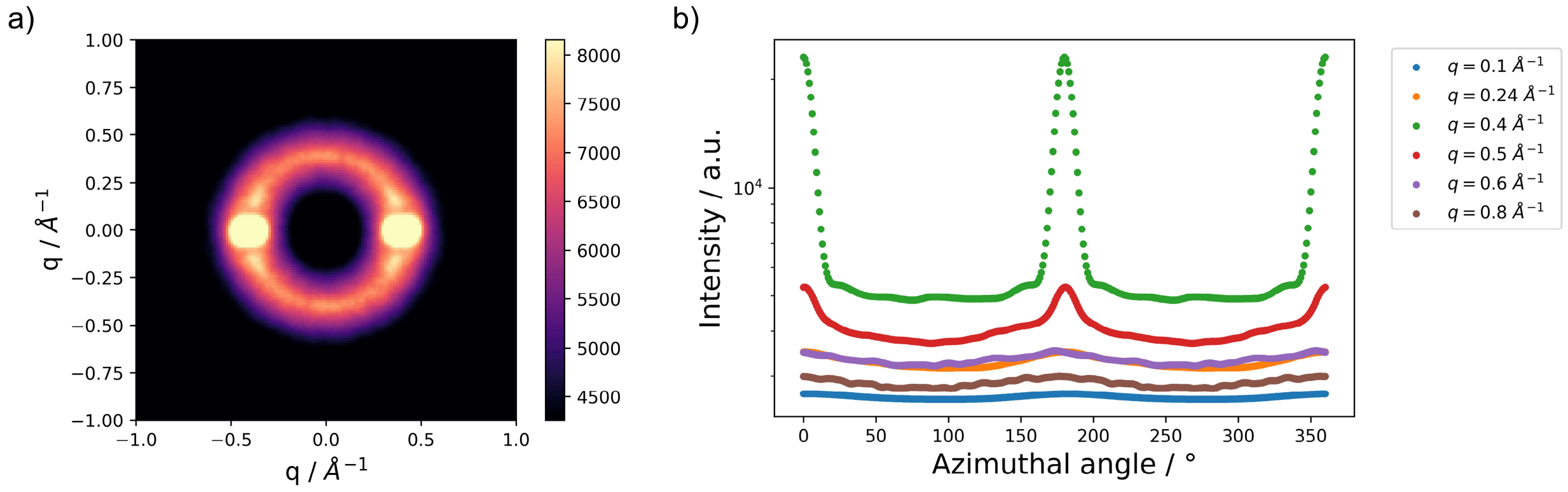

**Figure S5**. a) Median power spectrum of reoriented domains. b) Azimuthal lineouts of the median power spectrum for various ranges of reciprocal space (q) extracted from a). In-plane registration along the alkyl direction would be expected to produce diffraction or difusse scattering signal in the range $q \approx 0.2 - 0.3$ Å$^{-1}$ and approximately orthogonal to the common backbone reflection observed at $q = 0.4$ Å$^{-1}$ and azimuthal angle 0°.

## 3 HRTEM image processing

To locally map and quantify the predominant orientation of the backbone in IDT-BT thin-films, custom Python algorithms were developed based on the automated procedure reported by Takacs et al.[24,32] Briefly, the procedure consists of the following steps. Firstly, a bandpass filter centered at the spatial frequency of interest (i.e., 0.4 $Å^{-1}$ ± 0.06 $Å^{-1}$) is applied to the image stack, which had been previously multiplied by a 2D raised cosine window to reduce spectral leakage. A bandpass filter attenuates all spatial frequencies below and above the range of interest. Stacks of frames with apparent motion-induced image blurring are corrected with the method described below and a subset of image frames is selected to limit dosage to 10 – 15 $e^{-}Å^{-2}$ (fractionated over 8 – 12 frames). Secondly, the bandpass-filtered 2D projection is broken into overlapping (eight-fold oversampling) small tiles of 64 x 64 pixels (12.3 nm x 12.3 nm) or 128 x 128 pixels, and the local power spectrum of each tile is computed. Thirdly, a map of predominant backbone peak orientations is constructed by analyzing the local computed power spectra. As shown in Figure 1c in the main text, peaks on the local power spectrum indicate the presence of periodic features with d-spacing close to the backbone stacking distance. Peaks are considered to reflect the predominant orientation of the backbone at a spatial location on the image if their intensity is at least above two median absolute deviations from the background level. The background level was computed by imaging a region of the TEM grid that was not covered with the sample and applying the same procedure described above. Lastly, the extracted map of backbone peak orientations is utilized to compute the director fields and trace flow visualizations following guidelines by Takacs et al.[24,32] and Panova et al.[15] The map of backbone peak orientations is also an input variable for the local orientational spatial autocorrelations and domain identification algorithms.

## 4 Correction of motion-induced image blurring

A frequent occurrence in cryogenic HRTEM imaging is blurring due to the relative motion of the specimen and the holder. We can correct and measure concerted motion-induced blurring of HRTEM images by leveraging the high frame rate capabilities of direct electron detectors, which allow for the acquisition of short exposure, dose-fractionated stacks of frames on a particular region of the IDT-BT film. A widely used method to correct and measure spatial drift consists of performing a translation to align the features of interest in adjacent frames.[45,65,66] The concerted drift of the film relative to the holder can be tracked by computing the cross-correlation between the first and the other frames and measuring the relative displacement of the correlation peak from the origin. The cross-correlation approach assumes two images that are almost identical in content.

An example of the drift correction method is shown in Figure S6. Spatial drift on the starting 2D projected image stack can be identified in the frequency domain by different displacements along the spatial frequencies of interest in the FFT image. In the example in Figure S6, these displacements occur along the diagonal left of the FFT. To ensure the motion-correction algorithm identifies drift for the spatial frequencies of interest (i.e., backbone spacing), we first apply a bandpass filter centered at the spatial frequency range corresponding to the backbone repeat distance to every frame in the stack. The FFT of each bandpass-filtered frame is then computed to calculate the cross-correlation between the first and all other frames and measure the 2D relative displacement between frames. The frames are then realigned to correct for spatial drift. After drift correction, the FFT of the 2D projected image stack no longer presents displacements along the diagonal left, while preserving the backbone diffraction signal. We also note that after extended periods (>2.5 hours) to allow for mechanical stabilization of the sample before exposure, it was possible to record HRTEM images of IDT-BT that presented little or no spatial drift. We found no difference in the results of our analysis between HRTEM images that were or were not corrected for motion-induced blurring. We also note that while in the illustrative example in Figure S6 the drift correction method was applied to a stack of 48 frames (~ 60 $e^{-}Å^{-2}$), in this work we restricted electron beam dosage to 10 – 15 $e^{-}Å^{-2}$ and, consequently, we only applied drift correction methods to align the first 8 – 12 frames of the HRTEM movies.

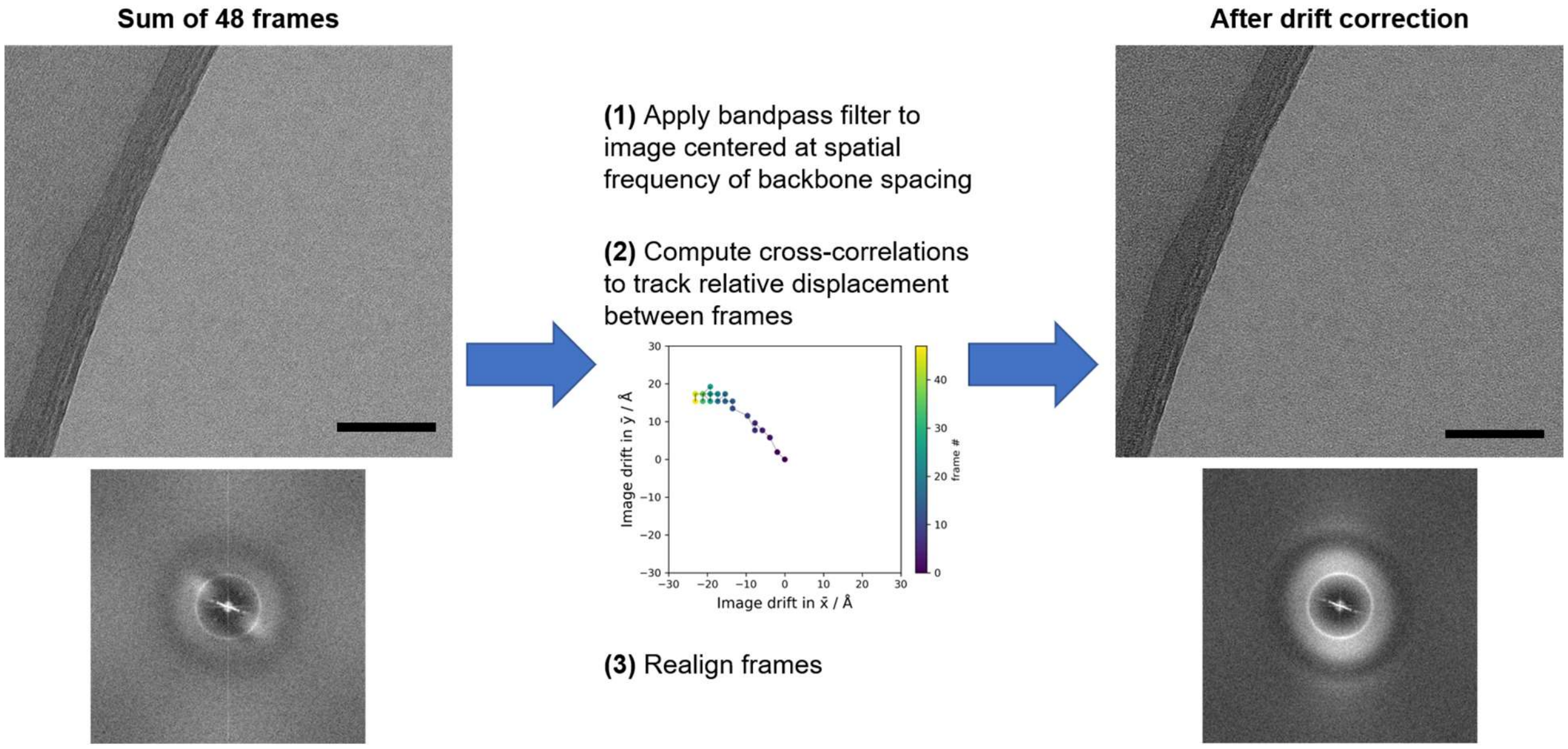

**Figure S6.** Example illustrating the correction of motion-induced image blurring in a HRTEM image with spatial drift. Scalebar is 50nm.

## 5 Domain identification and grain boundary analysis

Regions of aligned backbone were identified with in-house developed Python algorithms that spatially search for regions of alignment self-similarity. Briefly, we use the map of backbone peak orientations extracted by tiling the HRTEM images to search for adjacent regions that have backbone peak orientations within ± 5° of the median orientation of a forming cluster. The domain size is estimated as the square root of the total area of a domain. The minimum defined size for a domain is 5.4 nm, and smaller domains were not accounted for the domain size statistics. The algorithm accounts for the presence of multiple, overlapping backbone peaks present in the HRTEM image tiles, thus allowing for the identification of overlapping domains.

Over 130,000 domains were identified across several IDT-BT regions spanning an imaged area of ≈ 9 $\mu m^2$. As shown in Figure S7, all probed regions of the IDT-BT sample exhibit similar distributions. The median value of the domain size distribution is ≈ 10 nm and the interquartile range ($25^{th}$ to $75^{th}$ percentile) spans ≈ 7 – 16 nm. Outliers in the data sets (> 2.7 σ from the median) are represented with circles in Figure S7. It is important to note that given the minimum defined size for a domain is 5.4 nm, it is not possible to capture domains with smaller size and they are therefore not accounted for in the statistics. Grain boundary angles statistics were obtained by identifying the relative misorientation of each domain with its adjacent domains.

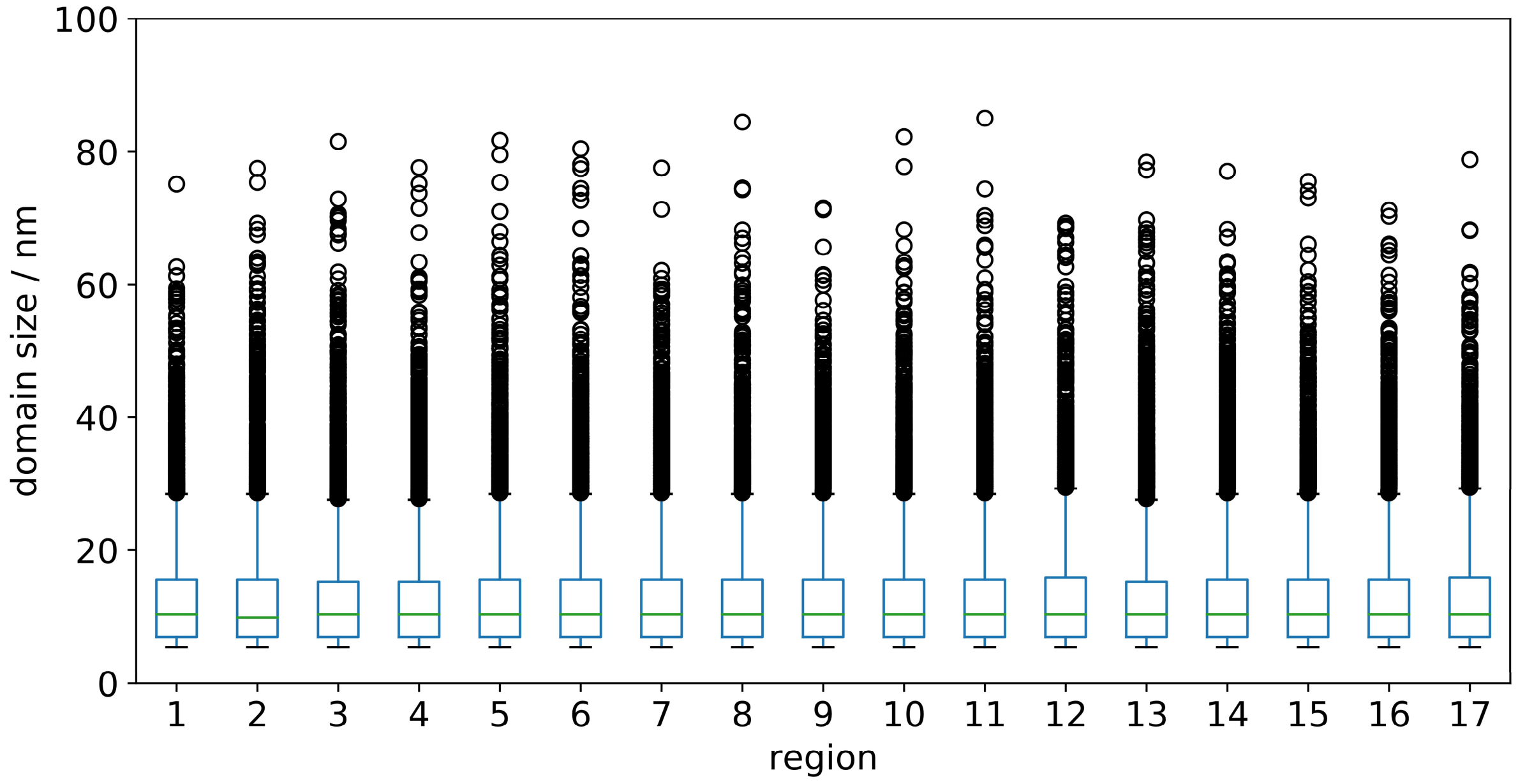

**Figure S7.** Domain size distribution across various regions of imaged IDT-BT.

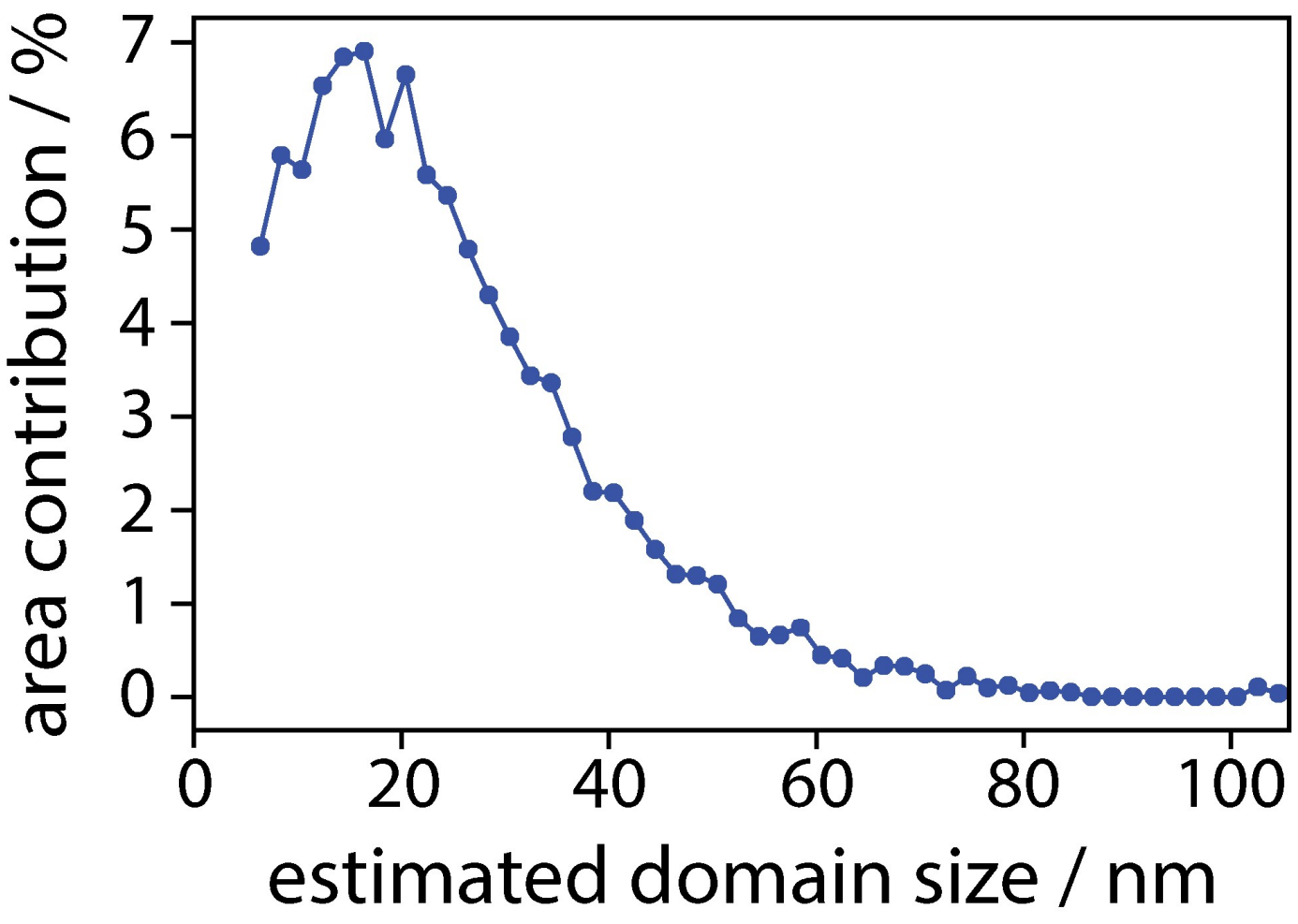

**Figure S8.** Area-weighted distribution of domain sizes calculated across various HRTEM images, equivalent to a probed area of ≈ 9 $\mu m^2$. The area-weighted distribution can be fit to a Gaussian distribution with mean ≈ 16 nm and peak width ≈ 30 nm.

## 6 Local orientational spatial autocorrelations

Autocorrelation functions are used to find patterns in data. We seek to find systematic variation between polymer regions as a function of two variables: relative distance (Δd) and misorientation (Δθ). Thus, we define the local orientational spatial autocorrelations as a metric to estimate the likelihood of finding two regions of the polymer thin-film separated by a distance Δd and presenting a misorientation Δθ. The starting point for computing the autocorrelations is the map of backbone peak orientations extracted, as described in Section 3, by tiling the HRTEM images. The autocorrelation function is computed by spatially and orientationally sorting the presence of two correlated backbone peaks:

$$C(\Delta d, \Delta\theta) = \frac{\sum f(d,\theta) \cdot f(d + \Delta d, \theta + \Delta\theta)|_{d,\theta}}{\sum f(d) \cdot f(d + \Delta d)|_{d}}$$

where $f(d, \theta)$ is a Boolean (i.e., 0 or 1) indicating whether at distance d and orientation θ a backbone peak is present (i.e., at distance d the polymer is predominantly oriented at angle θ). Similarly, $f(d + \Delta d, \theta + \Delta\theta)$ is a Boolean indicating whether there exists a backbone peak at a distance Δd with misorientation Δθ relative to the point defined in $f(d, \theta)$. The denominator term, $\sum f(d) \cdot f(d + \Delta d)|_{d}$, is the sum of backbone peaks at a relative distance Δd regardless of their orientation. This normalization term accounts for the fact that some distances are more likely to occur than others.

Figure S9a,b shows the computed autocorrelations for a simulated sample with random backbone orientations. The simulated sample is generated by creating a map of backbone peaks with random orientation, such that the average number of peaks at any (x, y) location in the sample is equivalent to the average number of peaks encountered at any (x, y) location in the IDT-BT images. The average number of peaks in IDT-BT was 2.7 peaks at any (x, y) out of 180 (i.e., θ = 0° to 179°) possible backbone peak orientations. The simulated sample of backbone orientations shows strong correlations at (Δd = 0 nm, Δθ = 0°) due to the correlation of the backbone peaks with themselves. The probability of two peaks being correlated at any other (Δd, Δθ) is random, and no preferential orientation trend is observed. In contrast, IDT-BT (Figure S10a,b) shows strong correlations within 5° for distances ~ 10 nm, suggesting polymer chains are likely to be aligned at this length scale. The positive correlations observed at low misorientation angles are emphasized by the low likelihood of backbone peaks to be misaligned between ~5 to 18° within ~ 10 nm.

The standard score (z-score) autocorrelation heatmaps and lineouts at various distances for IDT-BT and a simulated sample of random orientations are shown in Figure S10c,d and Figure S9c,d, respectively. The standard scores are calculated for each Δθ by subtracting the mean and dividing by the standard deviation of $C(\Delta d, \Delta\theta)$ for a given Δθ. Visualization of the standard scores instead of $C(\Delta d, \Delta\theta)$ ensures the patterns as a function of Δθ are not overwhelmed by patterns observed at other Δθ misorientation values where interactions are more recurrent (e.g., strong correlation for $\Delta\theta \lessapprox 5°$ due to domains of locally aligned chains). Positive standard scores in the autocorrelations indicate an enhancement in correlation, such that the likelihood of finding two aligned polymer regions separated by Δd for a particular value of Δθ is higher than the mean probability. Negative standard scores indicate a detriment in correlation, therefore the probability of finding two locally aligned polymer regions that are separated Δd at misorientation Δθ is less likely to occur than the mean probability.

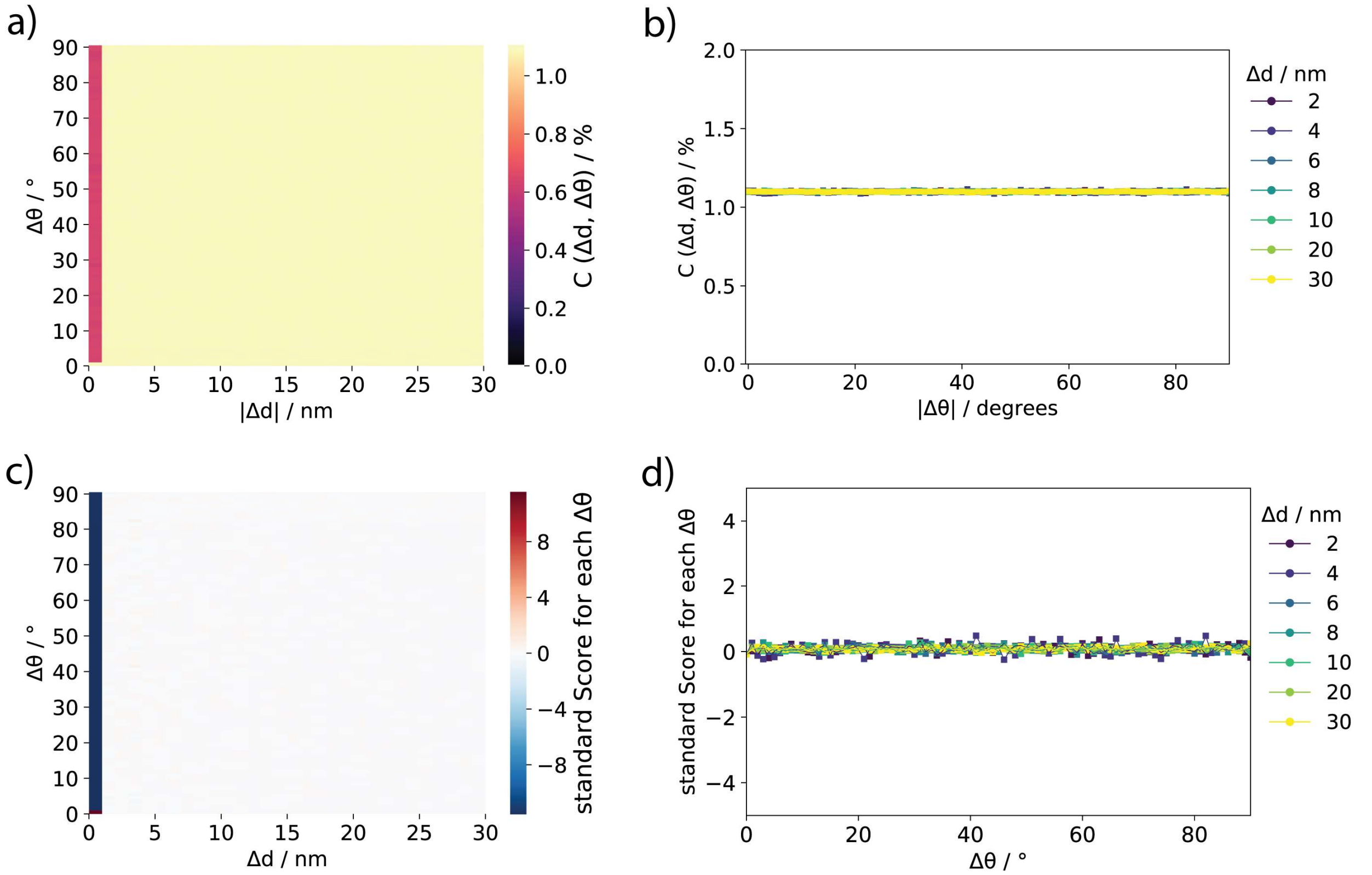

**Figure S9.** Local orientational spatial autocorrelations analysis for a simulated sample with random backbone orientations. **a)** Heatmap of the local orientational autocorrelations, $C(\Delta d, \Delta\theta)$. **b)** Linetcuts of $C(\Delta d, \Delta\theta)$ at selected distances distances $(\Delta d)$**. c)** Heatmap of the standard scores for the computed local orientational autocorrelations. **d)** Representative standard score lineouts at selected distances $(\Delta d)$.

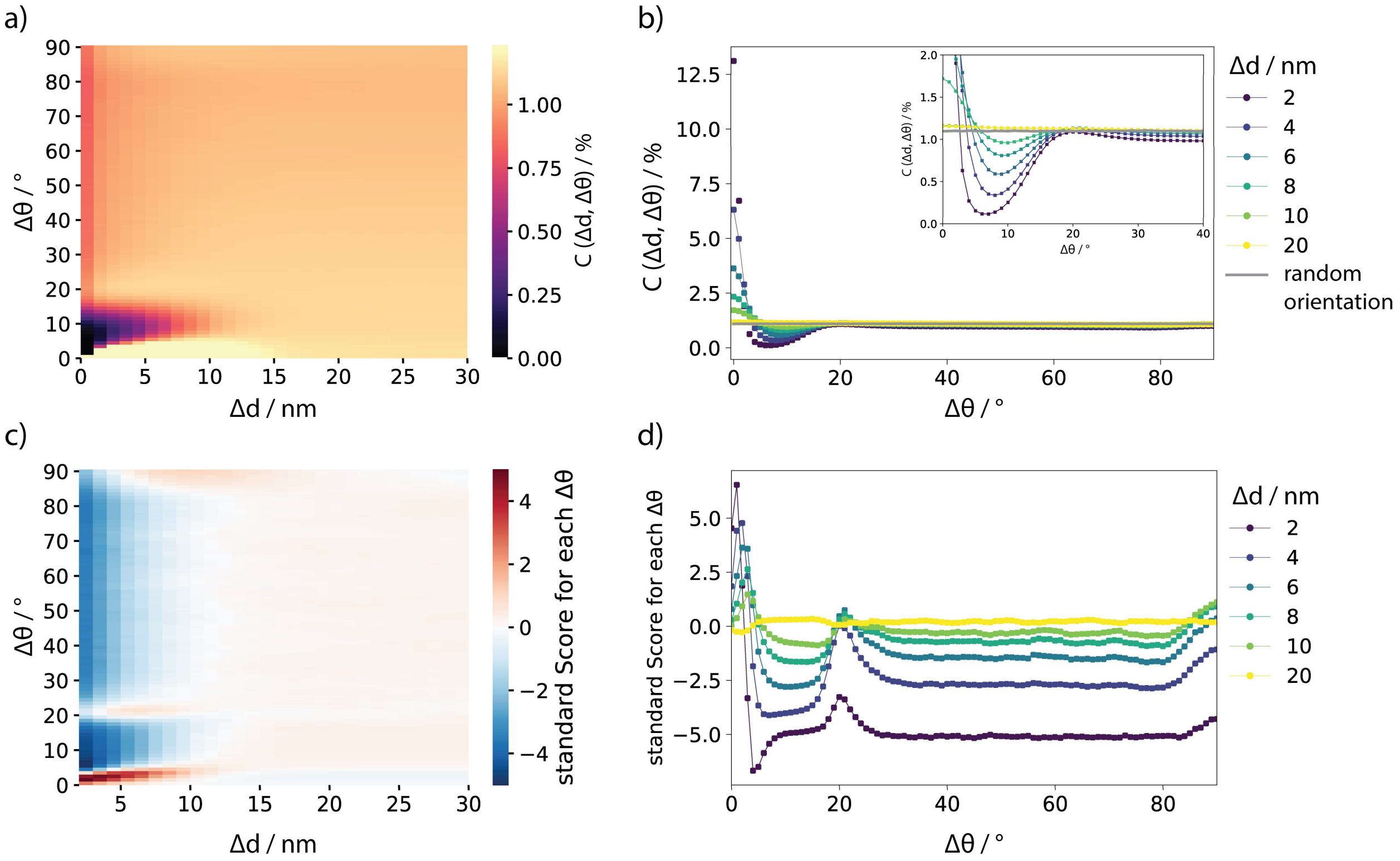

**Figure S10.** Local orientational spatial autocorrelations analysis for IDT-BT. **a)** Heatmap of the local orientational autocorrelations, $C(\Delta d, \Delta\theta)$. **b)** Linecuts of $C(\Delta d, \Delta\theta)$ at selected distances distances ($\Delta d$). A representative linecut (at $\Delta d = 20$ nm) for a simulated sample with random backbone orientations is shown in gray. **c)** Heatmap of the standard scores for the computed local orientational autocorrelations. **d)** Representative standard score lineouts at selected distances ($\Delta d$).